\documentclass[preprint,onecolumn,aps,pre,eqsecnum,a4paper]{revtex4-1}
\usepackage[latin9]{inputenc}
\usepackage{color}
\usepackage[british]{babel}
\usepackage{amsmath}
\usepackage{graphicx}
\usepackage[unicode=true,pdfusetitle,
 bookmarks=false,
 breaklinks=true,pdfborder={0 0 0},backref=false,colorlinks=true]
 {hyperref}

\makeatletter
\numberwithin{equation}{section}
\numberwithin{figure}{section}
\numberwithin{table}{section}

\DeclareGraphicsExtensions{.png,.pdf,.eps}
\graphicspath{{.}}

\makeatother

\begin{document}


\title{The super-indeterminism in orthodox quantum mechanics does not implicate
the reality of experimenter free will}

\author{Jan \surname{Walleczek}}

\email[Corresponding author: ]{walleczek@phenoscience.com}

\selectlanguage{british}%

\affiliation{Phenoscience Laboratories, Novalisstrasse 11, 10115 Berlin, Germany\vspace{1cm}
}
\begin{abstract}
The concept of \textquoteleft super-indeterminism\textquoteright{}
captures the notion that the \textit{free choice assumption} of orthodox
quantum mechanics necessitates \textit{only} the following requirement:
an agent\textquoteright s free-choice performance in the selection
of measurement settings must \textit{not} represent an exception to
the rule of irreducible \textit{quantum indeterminism} in the physical
universe (i.e, ``universal indeterminism''). Any \textit{additional}
metaphysical speculation, such as to whether quantum indeterminism,
i.e., intrinsic randomness, implicates the reality of experimenter
\textquotedblleft freedom\textquotedblright , \textquotedblleft free
will\textquotedblright , or \textquotedblleft free choice\textquotedblright ,
is redundant in relation to the predictive success of orthodox quantum
mechanics. Accordingly, super-indeterminism views as redundant also,
from a technical standpoint, whether an affirmative or a negative
answer is claimed in reference to universal indeterminism as a necessary
\textit{precondition} for experimenter freedom. Super-indeterminism
accounts, for example, for the circular reasoning which is implicit
in the free will theorem by Conway and Kochen~\citep{Conway.2006free,Conway.2009strong}.
The concept of super-indeterminism is of great assistance in clarifying
the often misunderstood meaning of the concept of \textquotedblleft free
variables\textquotedblright{} as used by John Bell~\citep{Bell.1977free}.
The present work argues that Bell sought an \textit{operational, effective
free will theorem}, one based upon the notion of \textquotedblleft determinism
\textit{without} predetermination\textquotedblright , i.e., one wherein
\textquotedblleft free variables\textquotedblright{} represent \textit{universally
uncomputable} variables. In conclusion, the standard interpretation
of quantum theory does not answer, and does not need to answer in
order to ensure the predictive success of \textit{orthodox} theory,
the question of whether either \textit{incompatibilism} or \textit{compatibilism}
is \textit{valid} in relation to free-will metaphysics and to the
\textit{free-will phenomenology} of experimenter agents in quantum
mechanics.
\end{abstract}
\maketitle

\section{Introduction}

\begin{figure}[!tbh]
\begin{centering}
\includegraphics[scale=0.48]{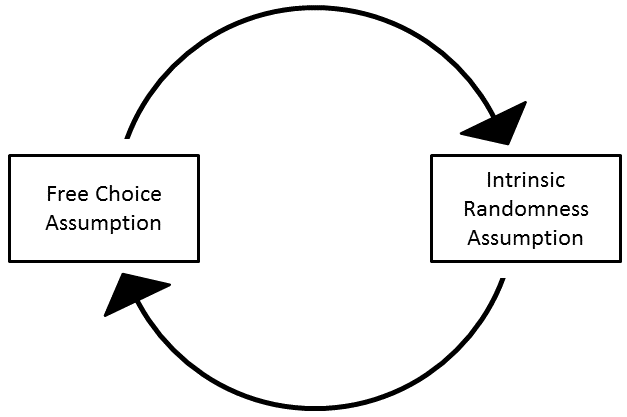}
\par\end{centering}

\caption{{\small{}Super-indeterminism in the orthodox interpretation of quantum
mechanics. Illustrated is the standard assumption that experimenter
}\textit{\small{}free choice}{\small{} depends on the existence of
a process demonstrating }\textit{\small{}intrinsic randomness}{\small{}
and }\textit{\small{}vice versa}{\small{}. Again, the reality of }\textit{\small{}universal
indeterminism}{\small{} depends equally -- in turn -- on experimenter
free will being }\textit{\small{}undetermined}{\small{}. For example,
the concept of \textquoteleft super-indeterminism\textquoteright{}
accounts for the much-cited, yet controversial, interpretation of
experimenter free will in orthodox quantum mechanics by Conway and
Kochen~\citep{Conway.2006free} who famously argued that \textquotedblleft if
indeed we humans have free will, then {[}so do{]} elementary particles.\textquotedblright \label{fig:1}}}
\end{figure}

The question concerning the reality of experimenter free will in science
in general, and in quantum mechanics in particular, has a long and
controversial history. As is well-known, the concept of free will,
as such, has eluded universal definition, no less in the context of
quantum theory and quantum indeterminism (e.g., Kane~\citep{Kane.2005contemporary,Kane.2011oxford}).
The present analysis will limit itself to the introduction -- into
the larger discussion on \textquotedblleft quantum free will\textquotedblright{}
-- of the concept of \textquotedblleft super-indeterminism\textquotedblright .
Its chief purpose is to draw attention to the fact that neither the
metaphysical position of \textquoteleft determinism\textquoteright{}
nor of \textquoteleft indeterminism\textquoteright{} is capable of
offering any decisive answers towards the reality of experimenter
free will in quantum mechanics; both, the \textit{compatibilist},
and \textit{incompatibilist}, interpretation of free will finds strong
support in the community of free-will philosophers (Kane~\citep{Kane.2005contemporary,Kane.2011oxford}).
For explanation, \textit{compatibilism} is the position that determinism
and free will are compatible, whereas \textit{incompatibilism} states
that they \textit{cannot} be. Again, the critical question of determinism
\textit{versus} indeterminism in quantum theory, such as, for example,
in the evaluation of the possibility of de Broglie--Bohm theory (Bohm~\citep{Bohm.1952interpr1,Bohm.1952interpr2})
need not be conflated with the \textit{additional} question of whether
determinism or indeterminism is consistent with, or even indicative
of, the reality of experimenter freedom. Fig.~\ref{fig:1} illustrates
the metaphysical position of \textquoteleft super-indeterminism\textquoteright .

\subsection{Physical definitions of experimenter free will: indeterminism \textit{versus}
determinism}

Regarding the free choice assumption, \textit{physical definitions}
in the literature on quantum foundations typically invoke concepts
such as the \textit{complete independence} from, or \textit{complete
lack of correlation} with, \textit{any} past physical events in nature.
A recent example of a concise definition of free choice in the context
of \textit{orthodox} quantum mechanics is the one by Colbeck and Renner~\citep{Colbeck.2013short}.
They suggest that a choice called \textquotedblright \dots{} \textit{A}
is free if the only variables it is correlated with are those it could
have caused.\textquotedblright{} By that definition, it is apparent
that a free choice action \textit{A} is one that (i) must be \textquotedblleft uncaused''
itself, and (ii) sets in motion an entirely new causal chain, literally
from \textquotedblleft nothing'', or at least starting from an assumption
of \textquotedblleft intrinsic randomness'' (compare Fig.~\ref{fig:1}).
The strict co-dependence of (non-causal) \textquoteleft intrinsic
randomness' and (uncaused) \textquoteleft experimenter freedom' in
the definition offered by Colbeck and Renner~\citep{Colbeck.2013short}
is matched, for example, by the metaphysical assumptions also underlying
the free will theorem by Conway and Kochen~\citep{Conway.2006free,Conway.2009strong}.
These latter authors summed up the metaphysical import of their (free
will) theorem by claiming that \textquotedblleft if indeed we humans
have free will, then {[}so do{]} elementary particles\textquotedblright{}
(see Fig.~\ref{fig:1}). It is significant for the present work that
Colbeck and Renner~\citep{Colbeck.2013short} conclude their own
efforts at defining \textit{physically} a free-will concept by referring
to the authority of \textit{John Bell}. They refer to Bell\textquoteright s
well-known, but often \textit{misinterpreted}, statement concerning
the meaning of \textquotedblleft free variables\textquotedblright{}
in the context of quantum-mechanical predictions (Bell~\citep{Bell.1977free}):
\textquotedblleft For me this means that the values of such variables
have implications only in their future light cones.\textquotedblright{}
In particular, Colbeck and Renner~\citep{Colbeck.2013short} have
implicated a correspondence between Bell\textquoteright s position
on experimenter freedom and their own incompatibilist (e.g., super-indeterministic)
position on free choice. In the following, the incongruence, however,
of these separate views will be demonstrated. The subsequent analysis
will demonstrate that John Bell did \textit{not} anticipate, and in
fact denied, any conflict between experimenter freedom and an entirely
\textit{deterministic} world view (Bell~\citep{Bell.1976theory,Bell.1977free,Bell.2004nouvelle}).

\section{John Bell\textquoteright s own \textquotedblleft free will theorem\textquotedblright{}
in quantum mechanics is compatibilist}

In contrast to both Conway and Kochen~\citep{Conway.2006free} and
Colbeck and Renner~\citep{Colbeck.2013short}, John Bell did \textit{not}
adopt an \textit{incompatibilist} approach to the problem of free
will in quantum mechanics, e.g., the super-indeterministic approach
illustrated in Fig.~\ref{fig:1}. Instead, Bell was a \textit{compatibilist}
when it came to the free choice assumption and the notion of freedom
in general. He argued consistently in the tradition of compatibilism
starting in 1976 (Bell~\citep{Bell.1976theory,Bell.1977free,Bell.2004nouvelle}).
Although Bell never presented his position on free will as a theorem
in name, he discussed in explicit terms the conceptual foundations,
including key metaphysical assumptions, which led him to propose a
free choice concept on an \textit{operational basis}, a position which
might be called \textquotedblleft effective indeterminism\textquotedblright,
or \textquotedblleft determinism \textit{without} predetermination\textquotedblright .
Specifically, Bell proposed that an experimenter agent serves in quantum
correlation experiments as a source of \textit{effectively} free variables,
i.e., of variables that are, according to Bell, \textquotedblleft \dots{}
effectively free for the purposes at hand\textquotedblright{} (Bell~\citep{Bell.1977free}).
\textquotedblleft For me this means\textquotedblright , Bell explained,
\textquotedblleft that the values of such variables have implications
only in their future light cones. They are in no sense a record of,
and do not give information about, what has gone before. In particular
they have no implications for the hidden variables v in the overlap
of the backward light cones\dots \textquotedblright{} (Bell~\citep{Bell.1977free}).
Contrary to the free choice concept proposed by Colbeck and Renner~\citep{Colbeck.2013short},
Bell\textquoteright s own approach to the free choice assumption is
\textit{not} founded upon intrinsic randomness or fundamental quantum
indeterminism, i.e., his approach does \textit{not} reflect \textit{super-indeterminism}
(see Fig.~\ref{fig:1}).

\subsection{John Bell\textquoteright s concept of \textquotedblleft free variables\textquotedblright{}
contradicts \textquotedblleft super-indeterminism\textquotedblright{}}

The fact that Bell was a compatibilist and that he thought of \textquoteleft freedom\textquoteright{}
as a concept that was \textit{not} in opposition to \textquoteleft determinism\textquoteright{}
is amply evident from his explanation in 1977 (Bell~\citep{Bell.1977free}).
Then, Bell admitted making an error in his original presentation of
1976 when he had wrongly proposed that \textquoteleft free\textquoteright{}
variables are \textquotedblleft \ldots{} not determined in the overlap
of the backward light cones\textquotedblright{} (Bell~\citep{Bell.1976theory}).
\textquotedblleft Here I must concede at once\textquotedblright ,
Bell~\citep{Bell.1977free} admitted \textquotedblleft that the hypothesis
becomes quite inadequate when weakened in this way. The theorem no
longer follows. I was mistaken.\textquotedblright{} Following this
admission, Bell affirmed that \textquoteleft free variables\textquoteright{}
are variables that are neither \textquoteleft not determined\textquoteright{}
nor \textquoteleft uncaused\textquoteright{} therefore, contradicting
the orthodox concept of super-indeterminism described in Fig.~\ref{fig:1}.
Instead, Bell\textquoteright s free variables are defined \textit{operationally},
as was mentioned already, as those that are \textquotedblleft \dots{}
effectively free for the purposes at hand\textquotedblright{} (Bell~\citep{Bell.1977free}).
Bell\textquoteright s crucial insight was this: a concept of \textit{effective}
experimenter free will must account for the fact that choices (actions)
of the experimenter agent are \textit{not pre-determined by the past
history of the universe}. In particular, the variables associated
with the choices by the epistemic agent must not be \textit{pre-determined},
although they will be \textit{unpredictably determined}, by \textquotedblleft the
hidden variables v in the overlap of the backward light cones\dots \textquotedblright{}
(Bell~\citep{Bell.1977free}). For the technical meaning of the notion
of \textquotedblleft unpredictably determined\textquotedblright{}
variables, in terms of \textquoteleft universally uncomputable' variables,
consult Sect.~4.

\subsection{John Bell\textquoteright s concept of \textquotedblleft free variables\textquotedblright{}
contradicts \textquotedblleft super-determinism\textquotedblright{}}

That Bell was well aware of the crucial distinction between, on the
one hand, \textquotedblleft \textit{pre}-determinism\textquotedblright ,
such as in the form of \textquotedblleft \textit{super}-determinism\textquotedblright ,
and, on the other hand, standard \textquotedblleft determinism\textquotedblright ,
is clearly evident from his final statements on the free will problem
(Bell~\citep{Bell.2004nouvelle}). \textquotedblleft One can envisage
theories\textquotedblright , Bell noted, \textquotedblleft in which
there just \textit{are} no free variables\ldots{} In such \textquoteleft super-deterministic'
theories the apparent free will of experimenters, and any other apparent
randomness, would be illusory.\textquotedblright{} Here, Bell notes
that in theories with \textit{unfree} variables, which he calls \textquotedblleft super-deterministic
theories\textquotedblright , experimenter free will is absent. Unfree
variables are those whose values are fully \textit{pre}-determined,
e.g., \textit{predictably} determined, by the past information in
the universe. The fact that Bell was a compatibilist and that he argued
for the possibility of \textit{effective} experimenter free will in
relation to a deterministic world view, is apparent from his clear
rejection of the \textit{super-deterministic} approach. \textquotedblleft I
do not expect to see a serious theory of this kind\textquotedblright ,
Bell~\citep{Bell.2004nouvelle} explained further. Instead, he proposed
that \textquotedblleft I would expect a serious theory to permit \textquoteleft deterministic
chaos' or \textquoteleft pseudorandomness', for complicated subsystems
(e.g., computers) which would provide variables sufficiently free
for the purpose at hand.\textquotedblright{}

\subsection{John Bell\textquoteright s concept of \textquotedblleft free variables\textquotedblright{}
implies emergent unpredictability }

The chief characteristic of \textquotedblleft deterministic chaos\textquotedblright ,
i.e., of the strictly \textit{nonlinear}, \textit{emergent} phenomenon
that is accounted for by \textit{self-organization} or \textit{complexity
theory}, is the \textit{uncertainty}, or \textit{unpredictability},
of observable outcomes. This is despite the fact that the chaotic
behavior is entirely governed by \textit{deterministic} relations.
In principle, this essential feature of deterministic chaos, holds
true for physical, chemical, biological, and even psycho-physical,
systems, including in neurophysiological brain states potentially
associated with the free-choice performance by an experimenter agent
(for an overview, e.g., Walleczek~\citep{Walleczek.2000self-organized}).
In short, the image used by Bell of \textquotedblleft deterministic
chaos\textquotedblright{} represents well the notion of \textquotedblleft effective
indeterminism\textquotedblright{} or \textquotedblleft determinism
\textit{without} predeterminism\textquotedblright . At the same time,
Bell\textquoteright s reference to \textit{deterministic chaos} and
\textit{pseudorandomness} is, of course, wholly \textit{incompatible}
with the intrinsic randomness assumption of orthodox quantum mechanics
shown in Fig.~\ref{fig:1}. Consequently, it is misleading to characterize
John Bell\textquoteright s own, well-documented views on free will
(Bell~\citep{Bell.1977free}) as being consistent with -- metaphysically
\textit{and} physically -- the \textit{incompatibilist} free-choice
concept described by Colbeck and Renner~\citep{Colbeck.2013short}. 

Finally, while the \textquotedblleft super-indeterminism\textquotedblright{}
in orthodox quantum mechanics has been considered before, although
not as a general metaphysical position (e.g., Hossenfelder~\citep{Hossenfelder.2012free}),
it was nevertheless argued by many others that a \textit{compatibilist
free will concept}, i.e., determinism, must be in violation of the
\textit{non-signalling constraint} (e.g., Barrett \textit{et al.}~\citep{Barrett.2005no};
Kofler \textit{et al.}~\citep{Kofler.2006experimenters}). The non-signalling
constraint mediates the \textit{apparent} conflict between quantum
theory and the theory of special relativity (e.g., Eberhard~\citep{Eberhard.1978bells}),
and its role in deciding the viability of \textit{compatibilist} free-will
concepts in quantum mechanics is reviewed next.

\section{The non-signalling theorem in the decision whether only an incompatibilist
free will concept is viable for quantum mechanics}

\begin{figure}[!h]
\begin{centering}
\vspace{3mm}
\includegraphics[scale=0.35]{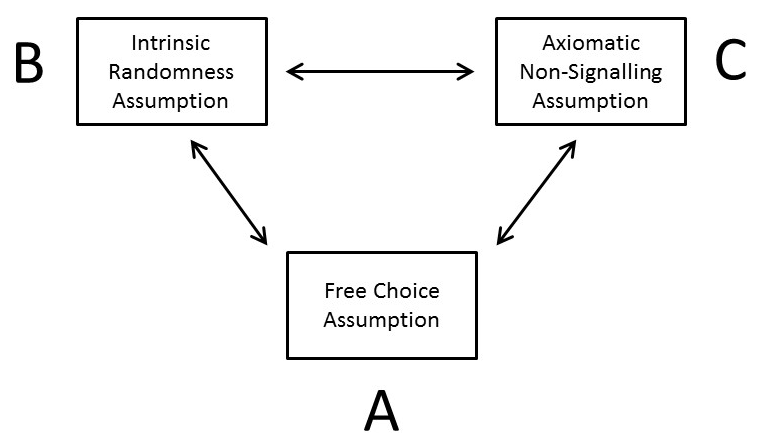}
\par\end{centering}

\caption{{\small{}An axiomatic non-signalling assumption does not represent
a conclusive argument for or against determinism and nonlocal hidden-variables
theories in quantum mechanics (from Walleczek and Grössing}~\citep{Walleczek.2016nonlocal}{\small{}).
As illustrated in Fig.}~\ref{fig:1}{\small{} also, there exists
-- in orthodox quantum mechanics -- an }\textit{\small{}interdependence}{\small{}
of free-choice (A) and intrinsic randomness (B) assumptions. The validity
of an }\textit{\small{}axiomatic}{\small{} non-signalling assumption
(C), a standard assumption in orthodox quantum mechanics, depends
in turn on the }\textit{\small{}independent}{\small{} validity of
assumptions A and B. Importantly, the concept of }\textit{\small{}super-indeterminism}{\small{}
(Fig.}~\ref{fig:1}{\small{}) demonstrates that an }\textit{\small{}independent}{\small{}
validation is out of reach however. This implies that the validity
of the }\textit{\small{}axiomatic}{\small{} interpretation of the
non-signalling theorem cannot be proven based on present knowledge
(for more details see Walleczek and Grössing}~\citep{Walleczek.2016nonlocal}{\small{}).\label{fig:2}}}
\end{figure}

A standard view holds that \textit{only} quantum theories invoking
\textit{universal} or \textit{fundamental indeterminism} could successfully
reproduce the predictions that are yielded by orthodox quantum mechanics.
Specifically when negating the involvement of \textquotedblleft cosmic
conspiracy\textquotedblright{} theories, such as \textit{super-determinism}
(e.g., Bell~\citep{Bell.2004nouvelle}), in the interpretation of
measurement outcomes, this view is widely thought to hold true. One
major reason is the suggested violation of the standard, i.e., axiomatic,
non-signalling constraint in any possible ontological quantum theory
(e.g., Colbeck and Renner~\citep{Colbeck.2011no,Colbeck.2012free};
Gallego \textit{et al.}~\citep{Gallego.2013full}). \textit{Axiomatic}
non-signalling, which was presumed in Bell\textquoteright s \textit{original}
interpretation of his theorem (Bell~\citep{Bell.1964einstein}),
is often considered mandatory, therefore, from the mainstream perspective
(see Fig.~\ref{fig:2}). However, as was argued recently, and is
illustrated in Fig.~\ref{fig:2}, the case for, or against, an axiomatic
non-signalling assumption is undecidable on logical grounds alone
(Walleczek and Grössing~\citep{Walleczek.2014non-signalling,Walleczek.2016nonlocal}).
The reason is the non-reducible, \textit{relational interdependence}
of the metaphysical assumptions that underlie such considerations
(see legend to Fig.~\ref{fig:2}).

\subsection{The two interpretations by John Bell of the non-signalling theorem}

\begin{figure}[!b]
\begin{centering}
\includegraphics[scale=0.35]{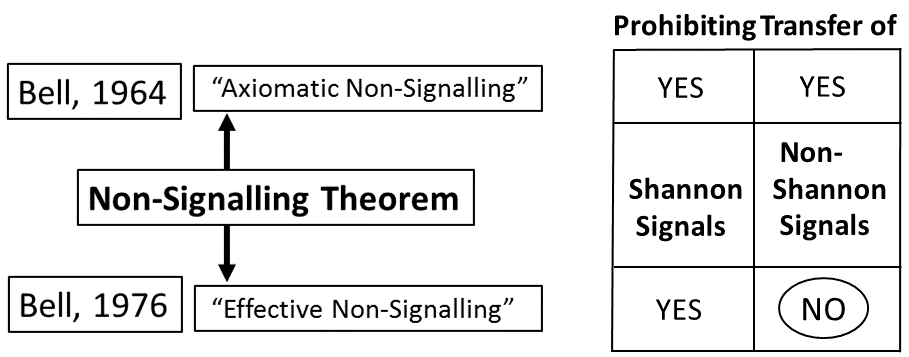}
\par\end{centering}

\caption{{\small The two different interpretations by John Bell of the non-signalling
theorem (from Walleczek and Grössing}~\citep{Walleczek.2016nonlocal}{\small{}).
Bell\textquoteright s first interpretation (Bell~\citep{Bell.1964einstein}),
here called \textquotedblleft axiomatic non-signalling\textquotedblright ,
rejects any form of deterministic quantum mechanics, upon assuming
experimenter freedom in the incompatibilist, super-indeterministic
sense (Fig.}~\ref{fig:1}{\small{}). The second, subsequent interpretation
(Bell~\citep{Bell.1976theory}), here called \textquotedblleft effective
non-signalling\textquotedblright , is compatible with ontological,
deterministic quantum approaches, including of the }\textit{\small{}nonlocal}{\small{}
type. Importantly, unlike axiomatic non-signalling, an effective non-signalling
theorem does }\textit{\small{}not}{\small{} prohibit the nonlocal
transfer of }\textit{\small{}uncontrollable}{\small{}, i.e., }\textit{\small{}non}{\small{}-Shannon,
signals. For definitions of Shannon and non-Shannon signals in the
context of Shannon\textquoteright s mathematical theory of communication
see Walleczek and Grössing}~\citep{Walleczek.2016nonlocal}{\small{}.
Effective non-signalling ensures the possibility of de Broglie--Bohm
theory, under an }\textit{\small{}effective free choice}{\small{}
assumption, in agreement with Bell\textquoteright s compatibilist
notion of \textquotedblleft free variables\textquotedblright{} (Bell~\citep{Bell.1977free}).\label{fig:3}}}
\end{figure}

As was reviewed recently, John Bell shifted his interest -- after
1976 -- from an \textquotedblleft axiomatic non-signalling\textquotedblright{}
assumption (Figs.~\ref{fig:2} and~\ref{fig:3}) towards an assumption
of \textquotedblleft effective non-signalling\textquotedblright{}
(Fig.~\ref{fig:3}), which is the non-signalling assumption that
is compatible with de Broglie--Bohm theory (Walleczek and Grössing~\citep{Walleczek.2016nonlocal}).
In the context of the present analysis, an \textit{effective} non-signalling
theorem \textit{only}, but not an \textit{axiomatic} one, is compatible
also with Bell\textquoteright s notion of \textquotedblleft free variables\textquotedblright{}
as described in Sect.~2. That is, an effective non-signalling theorem
allows for nonlocal quantum \textit{information transfers} while --
at the same time -- it prohibits \textit{superluminal signalling}
and communication. Fig.~\ref{fig:3} provides a summary (i) of the
two contrasting interpretations by John Bell of the non-signalling
theorem and (ii) of the \textit{communication-theoretic criterion}
which was shown to account for the difference between Bell\textquoteright s
original position of 1964 (Bell~\citep{Bell.1964einstein}) and his
later position which he first introduced in 1976 (Bell~\citep{Bell.1976theory}).

For the current analysis of the free will concept, the critical reason
for introducing an effective, instead of an axiomatic, non-signalling
condition is the following: an \textit{effective} non-signalling concept
eliminates the danger of contradiction between the free choice assumption
and the non-signalling principle, for example, in the context of de
Broglie--Bohm theory (Walleczek and Grössing~\citep{Walleczek.2016nonlocal}).
Next, given the different options for interpreting the notion of experimenter
freedom -- compatibilism \textit{versus} incompatibilism, Sect.~4
will offer an outlook towards the possible future of the free will
concept in quantum mechanics and beyond.

\section{The future of free will in quantum mechanics and beyond: How is freedom
possible?}

What is meant by the reality of experimenter free will? This work
defined \textit{the reality} of freedom -- until now -- simply as
the \textit{negation} of the statement that free will is an \textit{illusion}.
This preliminary (circular) definition drew on the earlier statement
by John Bell~\citep{Bell.2004nouvelle} who had explained that in
\textquotedblleft \dots{} \textquoteleft super-deterministic theories\textquoteright{}
the apparent free will of experimenters, and any other apparent randomness,
would be illusory.\textquotedblright{} Thus, Bell\textquoteright s
concept of super-determinism, as in \textquotedblleft super-pre-determinism\textquotedblright ,
implicates the complete absence of experimenter free will (see Sect.~2.2).
However, for those theories that do \textit{not} implicate super-determinism,
how could experimenter free will be possible? 

An informal, intuitive understanding of free will assumes that the
free-willed agent is \textquotedblleft free to choose\textquotedblright{}
among different options or alternatives. Also known as the principle
of alternative possibilities (e.g., Kane~\citep{Kane.2011oxford}),
the intuitive understanding instantly raises the problem of \textquotedblleft control\textquotedblright{}
in free-willed choices: Who or what is in charge when is performed
the \textquotedblleft intentional\textquotedblright{} selection of
\textit{one} option from a set of \textit{many} alternative options?
The concept of \textquoteleft experimenter \textit{freedom}\textquoteright{}
cannot, therefore, be reasonably discussed without considering also
the concept of \textquoteleft experimenter \textit{control}\textquoteright .
Indeed, generations of free-will philosophers have pointed out the
close connection between freedom \textit{and} control (for reviews
see Kane~\citep{Kane.2011oxford}). For example, free will would
be an illusion if an experimenter agent lacks (operational) control
over his or her performance, such as in setting up a \textit{certain}
experimental configuration for the purpose of \textquotedblleft asking
specific questions of nature\textquotedblright . An agent who can
serve as a \textquotedblleft source of operational control\textquotedblright{}
was previously defined as a \textit{type-1 agent}, whereas a \textquotedblleft source
of random variables\textquotedblright{} -- \textit{without} the element
of control -- was characterized as a \textit{type-2 source}. For explanation,
a \textquoteleft \textit{random event generator}\textquoteright{}
represents a \textit{type-2 source}, but it does \textit{not} possess
\textit{type-1 agency}, according to these definitions (Walleczek
and Grössing~\citep{Walleczek.2016nonlocal}).

\subsection{\textit{Quo Vadis} quantum mechanics? Dualism or monism? }

\begin{figure}[!tbh]
\begin{centering}
\includegraphics[scale=0.4]{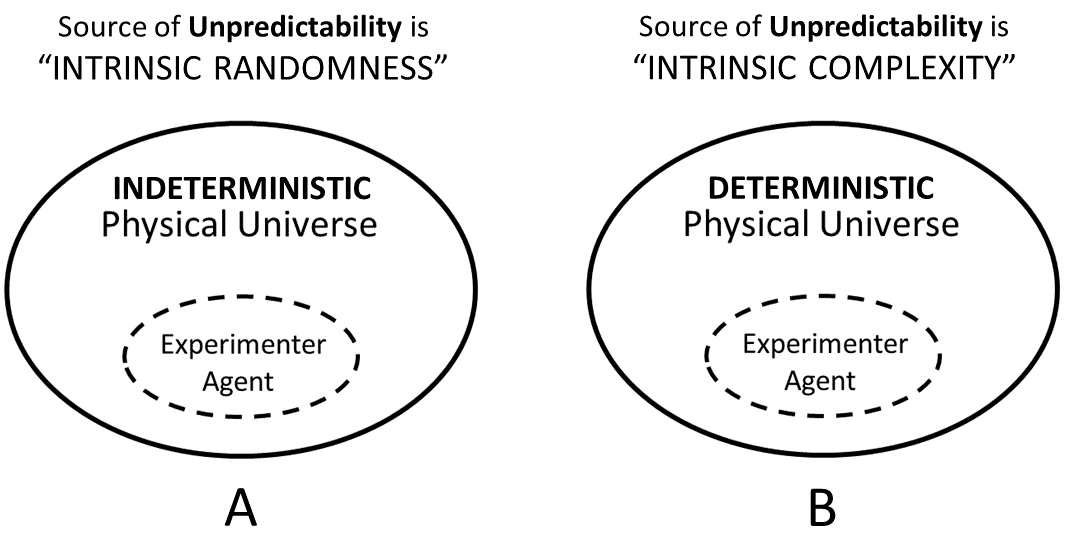}
\par\end{centering}

\caption{{\small{}Illustrating (A) }\textit{\small{}universal indeterminism}{\small{}
and (B) }\textit{\small{}universal determinism}{\small{}. (A) In \textquoteleft universal
indeterminism\textquoteright , agents and the physical universe are
subject to the same }\textit{\small{}fundamental indeterminism}{\small{}.
Here, the in-principle }\textit{\small{}unpredictability}{\small{}
of physical observables is a function of \textquotedblleft intrinsic
randomness\textquotedblright . Note that -- by itself -- indeterminism
neither rejects nor affirms the reality of \textquotedblleft free
will\textquotedblright{} or \textquotedblleft free choice\textquotedblright{}
as a result of }\textit{\small{}super}{\small{}-indeterminism (see
Fig.}~\ref{fig:1}{\small{}). (B) In \textquoteleft universal determinism\textquoteright ,
agents and the physical universe are subject to the same }\textit{\small{}fundamental
determinism}{\small{}. Here, the in-principle }\textit{\small{}unpredictability}{\small{}
of physical observables is a function of \textquotedblleft intrinsic
complexity\textquotedblright . Note that determinism need not be identified
with super-determinism in the form of super-}\textit{\small{}predeterminism}{\small{}
(see Sect.~2.2). This unnecessary, even false, identification of
determinism with pre-determinism would render any notions of \textquotedblleft free
will\textquotedblright{} or \textquotedblleft free choice\textquotedblright{}
}\textit{\small{}obsolete}{\small{} as part of scenario (B).\label{fig:4}}}
\end{figure}

Again, since \textquotedblleft randomness\textquotedblright{} is usually
viewed as the opposite of \textquotedblleft control\textquotedblright{}
it is hard to comprehend how \textquoteleft universal indeterminism\textquoteright{}
-- by itself -- could provide a realistic basis for the free-willed
actions of (type-1) agents. Similarly, \textquotedblleft non-randomness\textquotedblright ,
in the form of (pre)determinism, is equally unable to provide for
the capacity of \textit{controlling} physical events by an agent\textquoteright s
own free will. Consensus appears to be building towards the view that
neither determinism nor indeterminism provides an obvious pathway
towards freedom \textit{and} control (e.g., Kane~\citep{Kane.2005contemporary,Kane.2011oxford}).
Regarding the free-willed control by an agent of physical events in
nature, only an \textquotedblleft \textit{extra-physical factor}\textquotedblright{}
might advance the possibility of genuine freedom in the strong sense
of manifesting agent control. It certainly \textit{appears} to be
impossible to conceive of a scenario in which this extra-physical
\textquotedblleft causal power\textquotedblright{} (e.g., a concept
of \textquotedblleft free mental causation\textquotedblright{} by
an agent) would not be in violation of \textit{the} central tenet
of modern science: the rejection of \textit{dualism} in the form of
the \textit{Cartesian separation} of agent \textquotedblleft mind\textquotedblright{}
\textit{from} \textquotedblleft matter\textquotedblright{} (e.g.,
compare Cartesian \textquoteleft Res Cogitans\textquoteright{} \textit{versus}
\textquoteleft Res Extensae\textquoteright ). \textit{Quo Vadis} quantum
mechanics? \textit{Un-physical} dualism or \textit{physical} monism?
The final Fig.~\ref{fig:4} illustrates a comparison of the two competing
metaphysical frameworks that underlie the available options -- universal
indeterminism and determinism. What do these options mean for the
future of quantum mechanics, e.g., regarding the pressing issue of
\textit{operationalism versus realism} in quantum theory? 

Modern science rejects the Cartesian split, i.e., metaphysical \textit{dualism},
between agents and world. Instead, modern science posits that experimenter
agents are \textit{integral elements} (of reality) of the physical
universe, i.e., agents and universe -- together -- constitute a \textit{relational
physical continuum}. This \textit{non-dual} continuum principle is
indicated by the \textit{open} line in Fig.~\ref{fig:4} which encloses
the (operational) presence of the experimenter agent as an intrinsic
element of the universe. Importantly, natural science posits that
the behavior of both experimenter agent and the rest of the physical
universe is constrained by the same \textit{Laws of Nature}, e.g.,
quantum theory and relativity theory. This reflects the (monistic)
position of \textit{scientific physicalism}. Must the physicalist
position be \textit{radically revised} in light of \textit{orthodox}
quantum mechanics and of the concept of \textit{super-indeterminism}
in Fig.~\ref{fig:1}?

\subsection{Sources of unpredictability in universal indeterminism and determinism}

The above analysis suggests that the relationship between experimenter
freedom and experimenter control is neither addressed by universal
\textit{indeterminism} (Fig.~\ref{fig:4}a) nor by universal \textit{determinism}
(Fig.~\ref{fig:4}b). To repeat, the notion of \textquoteleft control\textquoteright{}
is usually identified with \textit{determination}, whereas the notion
of \textquoteleft freedom\textquoteright{} with \textit{indetermination}.
In any case, essential to any minimal conception of experimenter freedom,
as defined in the beginning of Sect.~4, appears to be the \textit{in-principle
unpredictability} of \textit{individual events} in the universe. Naturally,
this must include also an agent\textquoteright s free-choice performance
in the selection of measurement settings as an entirely unpredictable
event \textit{itself} (compare Fig.~\ref{fig:4}). On the one hand,
orthodox quantum mechanics offers the \textit{traditional} metaphysical
position in regards to the source of in-principle unpredictability:
\textquotedblleft \textit{intrinsic randomness}\textquotedblright{}
(Fig.~\ref{fig:4}a). On the other hand, concepts such as Bell\textquoteright s
notion of \textquotedblleft free variables\textquotedblright , for
example, offer an \textit{alternative} metaphysical position: \textquotedblleft \textit{intrinsic
complexity}\textquotedblright . This alternative position is consistent
with a \textit{universally} deterministic universe (see Fig.~\ref{fig:4}b).
Finally, Bell\textquoteright s free-variables concept will be briefly
discussed against the background of self-organization, complexity,
or emergence, theory which advances the notion of observables in nature
that are \textit{universally uncomputable}.

\subsection{John Bell\textquoteright s \textquotedblleft free variables\textquotedblright{}
as universally uncomputable variables }

Free variables are those that are \textquotedblleft \dots{} effectively
free for the purposes at hand\textquotedblright{} (Bell~\citep{Bell.1977free}).
As was reviewed in Sect.~2, in reference to ``free variables'',
John Bell raised the similarity to phenomena from nonlinear dynamics
such as \textquoteleft deterministic chaos', e.g., \textquoteleft pseudorandomness'.
The present work suggests that Bell\textquoteright s concept of free
variables can be justifiably defined as variables that \textit{emerge}
as a function of the \textit{intrinsic complexity} of universally
deterministic processes (compare Fig.~\ref{fig:4}b). These complex
and self-referential processes may generate physical observables whose
values are universally uncomputable, i.e., their computation would
require an \textit{infinite} amount of computational resources. In
this scenario, even \textit{Laplace's demon}, and \textquotedblleft \textit{the
universe itself}'', would be incapable of predicting with absolute
certainty the future outcome of \textit{individual} microscopic events
in nature. The concept of \textit{universal uncomputability}, which
is inherent in self-referential systems dynamics, plausibly offers
the kind of effective unpredictability which John Bell sought. For
example, see also the notions of ``effective indeterminism'' and
``determination without predetermination'' in Sect.~2.

\section{Conclusions}

The here introduced concept of \textit{super-indeterminism} for orthodox
quantum mechanics, which was shown to underlie the free will theorem
by Conway and Kochen~\citep{Conway.2006free,Conway.2009strong},
is consistent only with the metaphysical position shown in Fig.~\ref{fig:4}a:
\textit{intrinsic randomness}. By contrast, John Bell\textquoteright s
search for an \textit{effective free will} concept was shown to be
consistent only with the position shown in Fig.~\ref{fig:4}b: \textit{intrinsic
complexity}. It is likely that significant advances towards a \textit{realist}
quantum mechanics will depend on new insights into what might be called
also ``\textit{quantum complexity}'', i.e., complexity at the smallest
dimensions of reality. Summarizing, this work has identified two distinct
routes towards avoiding the ``cosmic pre-determination'' of the
free-choice performance by an experimenter agent: intrinsic randomness
and intrinsic complexity (Fig.~\ref{fig:4}). Importantly, neither
route was found to address the problem of the ``free-willed control\textquotedblright ~--
by an agent -- of any physical events in nature. Importantly, both
routes were argued to satisfy the minimal requirement of \textit{unpredictability},
i.e., lack of pre-determination, of an agent\textquoteright s choices
based upon past information about the universe. Finally, a better
understanding of the limits and possibilities of the alternative position
of \textit{quantum complexity} -- as a source of unpredictability~--
will require the development of a deeper understanding of the relationship
between quantum theory and the theory of complexity, self-organization,
and emergence. The prospects for the possibility of an \textit{emergent
quantum mechanics} critically depends on fashioning a deeper understanding
also -- in agreement with Bell\textquoteright s \textit{effective
free will} concept~-- of the \textit{relational} nature of an emergent
universe, including of emergent space and time. 
\begin{acknowledgments}
Work by Jan Walleczek at Phenoscience Laboratories (Berlin) is partially
funded by the Fetzer Franklin Fund of the John E. Fetzer Memorial
Trust. The author wishes to thank Gerhard Grössing and the research
group at the Austrian Institute for Nonlinear Studies, as well as
Nikolaus von Stillfried, for invaluable insights during many excellent
discussions.
\end{acknowledgments}

\providecommand{\href}[2]{#2}\begingroup\raggedright\endgroup

\end{document}